\documentclass[a4paper,11pt]{article}
\usepackage{pos}
\usepackage{slashed}     

\newcommand{\bra}[1]{\langle #1|}
\newcommand{\ket}[1]{|#1\rangle}

\newcommand{\di}{{\rm d}}
\newcommand{\ii}{i}
\newcommand{\e}{{\rm e}}

\def\wT{{\widehat T}}
\def\wj{{\widehat j}}
\def\wQ{{\widehat Q}}
\def\wP{{\widehat P}}
\def\wJ{{\widehat J}}
\def\wW{{\widehat W}}

\def\wrho{{\widehat{\rho}}}
\newcommand{\tr}{{\rm tr}}
\newcommand{\Tr}{{\rm Tr}} 
\newcommand{\Psibar}{{\overline \Psi}}

\title{Spin polarization in heavy-ion collisions induced by thermal vorticity and thermal shear}
\ShortTitle{Spin polarization in heavy-ion collisions}

\author*[a]{Matteo Buzzegoli}

\affiliation[a]{Department of Physics and Astronomy, Iowa State University,\\
  2323 Osborn Drive, Ames, Iowa 50011, USA}

\emailAdd{mbuzz@iastate.edu}

\abstract{
The vorticity is a quantity defined in a relativistic fluid that describes how much a fluid element is rotating and is accelerating. 
By measuring the spin polarization of hadrons, it was found that the quark gluon plasma produced in heavy-ion collisions is the most ``vorticous'' fluid ever observed. More generally, this opens the possibility to study the physics of QCD matter using spin.
Here I use statistical quantum field theory applied to a fluid in  local thermodynamic equilibrium to show how to connect the average spin of a fermion with hydrodynamic quantities, and in particular with the thermal vorticity and the thermal shear.
I show that the spin polarization of a Dirac particle induced by thermal vorticity is related to the gravitational (in medium) form factor related to spin-rotation coupling.
For these reasons, as we are understanding the role of spin in hydrodynamics and in heavy-ion collisions, spin is becoming a promising tool to investigate the properties of QCD and whose applications are just begun to be explored.}

\FullConference{25th International Spin Physics Symposium (SPIN 2023)\\
 24-29 September 2023\\
 Durham, NC, USA\\}

\begin{document}
\maketitle

\section{Introduction}
One of the goals of relativistic heavy-ion collisions is to study the physics of finite temperature QCD.
By measuring the momentum spectrum of particles and comparing with the theoretical predictions of momentum distributions and correlations, we can infer what happens during a collision.
Most notably, in this way it was found that the nuclear system goes through a phase of matter named quark gluon plasma (QGP) that behaves like a nearly perfect fluid.
Before the measurement of spin polarization of $\Lambda$ particles in 2017 \cite{STAR:2017ckg}, all the studies were based on the observation of momentum.

This paradigm started to change when it was realized \cite{Liang:2004ph,Voloshin:2004ha} that off central collisions should generate a large angular momentum and that particles, due to the spin-orbit coupling, should have their spin aligned along the direction of the total orbital angular momentum.
Later, describing the spin degrees of freedom as close to the local thermal equilibrium (in the same fashion as successfully done for the momenta), the spin polarization was connected to the collective motion of the plasma, and precisely to the thermal vorticity \cite{Becattini:2007sr,Becattini:2007nd,Becattini:2013fla}.
The observation of global spin polarization of $\Lambda$s \cite{STAR:2017ckg} is in good agreement with the hydrodynamic predictions, for more detail see for instance the review~\cite{Becattini:2020ngo}.

More recently, the measurement of the global polarization of $\Xi$ hyperons \cite{STAR:2020xbm} further confirmed the hydrodynamic picture since it showed that spin polarization in heavy-ion collisions, unlike $pp$ collisions, is not sensitive to specific hadron properties.
Over the past decade there has been a significant development of spin physics for heavy-ion collisions,
and spin polarization has also been measured at very high \cite{ALICE:2019onw,ALICE:2021pzu} and at very low \cite{STAR:2021beb,Kornas:2020qzi} energy.

The first measurement of spin polarization as a function of momentum, referred to as local spin polarization, was reported in \cite{STAR:2019erd} and is in disagreement with the predictions obtained from thermal vorticity. As I will show below, the agreement with the hydrodynamic picture is restored once it is realized that also other collective motions, in particular the shear flow, can induce a spin polarization. I will show how to use quantum statistical field theory to derive the connections between the spin polarization and the hydrodynamic quantities, and I will discuss the features of the shear induced polarization. I will then connect the vorticity induced polarization with the spin-orbit coupling and the related form factor.

\section{Spin polarization in quantum statistical field theory}
The spin of a massive relativistic particle in quantum mechanics is defined through the Pauli-Lubanski operator
\begin{equation}\label{luba}
 \widehat S^\mu = -\frac{1}{2m} \epsilon^{\mu \nu \rho \sigma} \wJ_{\nu \rho} 
 \widehat P_\sigma ,
\end{equation} 
where $\wJ$ is the angular momentum-boost operator and $\wP$ the energy-momentum operator.
The spin polarization measured in heavy-ion collisions is the average of the Pauli-Lubanski operator for a particle of given momentum $p$, denoted by $S^\mu(p)$, 
and it is obtained as the mean value of the operator (\ref{luba}) with the density matrix operator $\wrho$, that is
\begin{equation}\label{spinvector}
  S^\mu(p) = \Tr (\, \wrho \, \widehat S^\mu(p) \,) .
\end{equation} 
It is convenient to relate the average spin vector defined in Eq. (\ref{spinvector}) with the covariant Wigner functions,
which, for a the Dirac field  $\Psi$, are defined as
\begin{eqnarray}\label{wigdirop}
W_{\alpha\beta}(x,p) 
   &=& \int  \frac{\di^4 y}{(2\pi)^4} \; \e^{-\ii p \cdot y} \langle: 
   \Psibar_\beta \left(x+\frac y2\right) \Psi_\alpha \left(x-\frac y2\right):\rangle  \,,
\end{eqnarray}
where $\alpha$ and $\beta$ are spinorial indices, $:\; :$ denotes the normal ordering of creation and destruction operators, and angle bracket $\langle \: \rangle$ denotes the trace with the density matrix operator. When the momentum $p$ is future time-like, the Wigner functions contains only the contribution form the particle part of the Dirac field, and the following traces
\begin{equation}
\label{eq:WignFA}
\mathcal{F}_+(x,p) = \tr\left[ \, W(x,p) \theta(p^2)\theta(p^0)\right] , \quad 
\mathcal{A}_+^\mu(x,p) = \tr\left[ \gamma^\mu\gamma^5 W(x,p) \theta(p^2)\theta(p^0)\right],
\end{equation}
are the particle part of the  scalar and axial Wigner functions.
For a weakly interacting field, it can be shown \cite{Becattini:2020sww} that the average spin vector can be evaluated from the integrals
of $\mathcal{F}_+$ and $\mathcal{A}_+^\mu$ over a space-like hyper-surface $\Sigma$ as follows
\begin{equation}\label{meanspf0}
S^\mu(p) = \frac{1}{2} \frac{\int_\Sigma \di \Sigma \cdot p \;\mathcal{A}_+^\mu(x,p)}{\int_\Sigma \di \Sigma \cdot p \;\mathcal{F}_+(x,p)} .
\end{equation}
Predictions for the spin polarization can be obtained from this formula if the statistical ensemble, described by a density matrix $\wrho$, is known.
 
The covariant density matrix at the hypersurface $\Sigma$ is derived assuming that the system reached the local thermodynamic equilibrium at an initial stage and maximizing the entropy of the system.
Neglecting the dissipative effects, this procedure yields \cite{Becattini:2019dxo}
\begin{equation}\label{densopLE}
  \wrho_{\rm LE} = \frac{1}{Z} 
  \exp \left[ -\int_{\Sigma} \di \Sigma_\mu(y) \left( \wT^{\mu\nu}(y) 
  \beta_\nu(y) - \zeta(y) \wj^\mu(y) \right) \right],
\end{equation}
with $\wT$ the energy momentum tensor, $\wj$ a conserved current, $\beta=u/T$ is the inverse four temperature vector and $\zeta=\mu/T$ where $\mu$ is the chemical potential.
This operator is needed to obtain the Wigner functions (\ref{eq:WignFA}) at the point $x$.
Since the interaction scales are shorter than the macroscopic ones, the Wigner functions (\ref{eq:WignFA}) can be obtained
expanding the hydrodynamic fields in the operator (\ref{densopLE}) around the same point $x$ where they are to be evaluated, that is, setting $\zeta=0$ for simplicity,
\begin{equation}
	\label{eq:TaylorBeta}
	\beta_\nu(y) \simeq \beta_\nu(x) + \partial_\lambda \beta_\nu(x) (y-x)^\lambda.
\end{equation}
The resulting density matrix is
\begin{equation}
\label{eq:ApproxHydro}
\widehat{\rho}_{\rm LE} \simeq \frac{1}{Z} \exp\left[ - \beta_\nu(x) \widehat{P}^\nu 
		+ \frac{1}{2} \varpi_{\mu\nu}(x) \widehat{J}^{\mu\nu}_x  -\frac{1}{2} \xi_{\mu\nu}(x) \widehat{Q}^{\mu\nu}_x
		+\cdots\right]\,,
\end{equation}
where the antisymmetric derivative of $\beta$ is the thermal vorticity
\begin{equation}\label{thvort}
  \varpi_{\mu\nu} = -\frac{1}{2} \left( \partial_\mu \beta_\nu - \partial_\nu \beta_\mu \right),
\end{equation}
and the symmetric one is the thermal shear
\begin{equation}
	\xi_{\mu\nu}=\frac{1}{2}\left(\partial_\mu\beta_\nu + \partial_\nu\beta_\mu \right).
\end{equation}
Thermal vorticity couples with the conserved angular momentum operator
\begin{equation}
\widehat{J}^{\mu\nu}_x =\! \int \!{\rm d} \Sigma_\lambda \left[ (y-x)^\mu \widehat{T}^{\lambda\nu}(y) -
	(y-x)^\nu \widehat{T}^{\lambda\mu}(y)\right],
\end{equation}
while the thermal shear couples with a non-conserved symmetric quadrupole like operator
\begin{equation}
\widehat{Q}^{\mu\nu}_x =\! \int_{\Sigma_{FO}} \!{\rm d} \Sigma_\lambda \left[ (y-x)^\mu \widehat{T}^{\lambda\nu}(y) + 
	(y-x)^\nu \widehat{T}^{\lambda\mu}(y)\right].
\end{equation}
The definition of thermal vorticity (\ref{thvort}) implies that $\varpi$ is related to the acceleration of the fluid $a^\mu=u\cdot\partial u^\mu$, the relativistic angular velocity $\omega^\mu=\frac{1}{2}\epsilon^{\mu\nu\rho\sigma}u_\sigma \partial_\nu u_\rho$ and to the temperature $T=1/\sqrt{\beta^2}$ as follows
\begin{equation}
\varpi^{\mu\nu}=-\frac{1}{2}\left[\partial^\mu\left(\frac{1}{T}\right)u^\nu-\partial^\nu\left(\frac{1}{T}\right)u^\mu\right]
	+\epsilon^{\mu\nu\rho\sigma}\frac{\omega_\rho}{T}u_\sigma + \frac{1}{2T}\left(a^\mu u^\nu - a^\nu u^\mu\right) \, .
\end{equation}
While the thermal shear is expressed as
\begin{equation}
\xi^{\mu\nu}=\frac{1}{2}\left[\partial^\mu\left(\frac{1}{T}\right)u^\nu+\partial^\nu\left(\frac{1}{T}\right)u^\mu\right]
	+ \frac{1}{2T}\left(a^\mu u^\nu + a^\nu u^\mu\right) + \frac{1}{T}\sigma^{\mu\nu} + \frac{1}{3T}\theta\Delta^{\mu\nu} \, ,
\end{equation}
where $\nabla_\mu=\partial_\mu -u_\mu(u\cdot\partial)$, $\Delta^{\mu\nu}=\eta^{\mu\nu}-u^\mu u^\nu$, $\theta=\nabla\cdot u$ and $\sigma$ is the shear tensor
\begin{equation}
\sigma^{\mu\nu} = \frac{1}{2}\left(\nabla^\mu u^\nu + \nabla^\nu u^\mu\right) -\frac{1}{3}\Delta^{\mu\nu}\theta\, .
\end{equation}

Using the statistical operator in Eq. (\ref{eq:ApproxHydro}) and linear response theory, the average of the spin vector (\ref{meanspf0})
for a free Dirac field result in \cite{Becattini:2021suc}
\begin{equation}
\label{eq:SpinPolFirstOrder}
S^\mu(p) = - \frac{\epsilon^{\mu\rho\sigma\tau} p_\tau}{8m \int_{\Sigma} {\rm d} \Sigma \cdot p \; n_F}  
\int_{\Sigma} {\rm d} \Sigma \cdot p \;
n_F (1 -n_F) \left[ \varpi_{\rho\sigma} + 2\hat t_\rho \frac{p^\lambda}{E_p} \xi_{\lambda\sigma}\right] \,, 
\end{equation}
where $\hat{t}$ is the time direction in the laboratory frame and $n_F$ is the Fermi-Dirac phase-space distribution function:
\begin{equation}
	n_F = \frac{1}{\exp[\beta\cdot p - \mu q]+1},
\end{equation}
where $q$ is the charge of the particle.

This recently found \cite{Liu:2021uhn,Becattini:2021suc} contribution of spin polarization from thermal shear does not affect the predictions of global spin polarization that are well reproduced by thermal vorticity alone, but it is the dominant contribution in local spin polarization \cite{Fu:2021pok,Becattini:2021iol}. Quantitative agreement with the data of local spin polarization has been found if hadronization occurs at nearly constant temperature \cite{Becattini:2021iol}, as expected at the energies where data has been taken.
Further analysis of the impact of thermal shear on spin polarization \cite{Ryu:2021lnx,Alzhrani:2022dpi,Wu:2022mkr} revealed a significant dependence on the initial conditions of the hydrodynamic model.
As the theory for spin polarization does not require any additional parameter compared to the standard model of heavy-ion collisions, the measurements of local spin polarization have the potential to further constraint the initial conditions of the hydrodynamic equations and to probe the properties of the QGP, such as the shear and bulk viscosities.

\section{In medium form factors}
As mentioned, the spin polarization can be understood as a result of spin-rotation coupling.
Stating that a medium is rotating is equivalent of saying that there is an effective force causing an acceleration resulting in the circular motion.
The interaction of a single constituent of the medium with the overall rotation can then be studied as an inertial effect caused by a rotating observer.
The analysis of the Dirac equation in rotating coordinates revealed \cite{deOliveira:1962apw,Hehl:1990nf} that a Dirac particle in a medium rotating with angular velocity $\vec\Omega$ has the Hamiltonian
\begin{equation}
\label{eq:RotatingH}
H = H_0 - \vec{J}\cdot\vec{\Omega},
\end{equation}
where $\vec{J}=\vec{L}+\vec{S}$ is the total angular momentum of the fermion and $H_0$ is the Hamiltonian in absence of rotation.
This is analogous to the interaction of a fermion with an external magnetic field, resulting in the Zeeman effect.
It follows that the energy is lower when the spin is aligned along the rotation, hence spin is polarized along the rotation when the system is close to equilibrium.

More generally, we can describe the interaction with weak gravitational field through the Lagrangian
\begin{equation}
\mathcal{L}_{int} = \frac{1}{2}(g_{\mu\nu}-\eta_{\mu\nu})T^{\mu\nu},
\end{equation}
which reproduces the Hamiltonian (\ref{eq:RotatingH}) if we choose $g$ as the metric of rotating coordinates.
Using scattering theory, the coupling between spin and rotation can be studied looking at the matrix elements of the energy momentum tensor,
that can be decomposed in terms of the form factors $f_1$ and $g_\Omega$ as follows
\begin{equation}
\langle p',\,s'| \widehat{T}^{\mu\nu}(0)|p,\,s\rangle = \bar{u}(p',s')\left[ f_1(q^2) \frac{P^\mu P^\nu}{m}+g_\Omega(q^2)\frac{\sigma^{(\mu\alpha}q_\alpha P^{\nu)}}{2m}
	+ O(q^2)\right] u(p,s),
\end{equation}
where $\ket{p,\, s}$ is a one particle state with momentum $p$ and spin $s$, $u(p,s)$ and $\bar{u}(p',s')$ are spinors, $q=p'-p$ and $P=(p+p')/2$.
The form factor $g_\Omega$ is also called the gravitomagnetic moment and it can be shown \cite{Buzzegoli:2021jeh} that describes exactly the spin rotation coupling in (\ref{eq:RotatingH}).
It has long been known that, differently form other form factors like the magnetic moment, $g_\Omega$ can not receive radiative corrections and should always be $g_\Omega=1$ otherwise the theory is not Lorentz covariant and would violate the Einstein Equivalence Principle (EEP) \cite{Kobzarev:1962wt,Cho:1976de,Teryaev:2016edw}.

This restriction is lifted when considering a system in thermal equilibrium. Indeed, the presence of a thermal bath breaks the Lorentz invariance and as a result finite temperature field theory violates the EEP  \cite{Donoghue:1984zs,Donoghue:1984ga}. As a consequence, the matrix elements of an operator in a medium can have additional structure and additional form factors \cite{Buzzegoli:2021jeh,Lin:2023ass}. In addition to four-vectors $P$ and $q$, matrix elements can also be given in terms of $u$, that denotes the fluid velocity, and the vector $l^\mu=\epsilon^{\mu\nu\rho\sigma}u_\nu P_\rho q_\sigma$. The energy-momentum tensor is then decomposed as
\begin{equation}
\label{eq:EMTinMedium}
\begin{split}
\bra{p',s'}\widehat{T}_{\mu\nu}(0)&\ket{p,s}=\bar{u}(p',s')\Big\{I_{P\gamma}(P,q)\left(P_\mu \gamma_\nu + P_\nu \gamma_\mu\right)
		+I_{u\gamma}(P,q)\left(u_\mu \gamma_\nu + u_\nu \gamma_\mu\right)+\\
		&+I_{Pl}(P,q)\slashed{\hat{l}}\left(P_\mu \hat{l}_\nu + P_\nu \hat{l}_\mu\right)
		+I_{ul}(P,q)\slashed{\hat{l}}\left(u_\mu \hat{l}_\nu + u_\nu \hat{l}_\mu\right)\Big\}u(p,s)+\cdots,
\end{split}
\end{equation}
where $\hat{l}^\mu=l^\mu/\sqrt{-l^2}$ and the in-medium spin-rotation coupling is \cite{Buzzegoli:2021jeh}
\begin{equation}
g_\Omega(P,q)=4\left(I_{P\gamma}(P,q) + \frac{I_{u\gamma}(P,q)}{(P\cdot u)} -I_{Pl}(P,q) -\frac{I_{ul}(P,q)}{(P\cdot u)} \right).
\end{equation}
Explicit calculations in finite temperature QED \cite{Buzzegoli:2021jeh}, revealed that indeed the gravitomagnetic moment receives radiative corrections,
for instance, in the low temperature limit $T \ll m$, the 1-loop gravitomagnetic moment is
\begin{equation}
\label{eq:AGMQED}
\lim_{\substack{q\to 0 \\ P^2\to (P\cdot u)^2}} g_\Omega= 1 -\frac{1}{6}\frac{e^2 T^2}{m^2}.
\end{equation}

In \cite{Buzzegoli:2021jeh} it was shown that $g_\Omega$ is directly connected to the axial vortical effect, that is the generation of an axial current induced by the rotation of the medium.
Here I use the same method to show how spin-rotation coupling is related to the spin polarization (\ref{spinvector}).
In order to apply the formula (\ref{meanspf0}), I first need to compute the Wigner function in a system in thermal equilibrium with thermal vorticity.
Such a system is described by the density matrix in Eq. (\ref{densopLE}).
The contribution of thermal vorticity $\varpi$ to the Wigner functions (\ref{wigdirop}), obtained using linear response theory to the density matrix in Eq. (\ref{densopLE}), is
\begin{equation}\label{eq:DeltaVortWig}
\Delta_\varpi W_{ab}^+(x,k) = -\int_0^1\di z\int_\Sigma \di\Sigma_\lambda(y)\varpi_{\rho\kappa}(y-x)^\kappa\langle \wW_{ab}^+(x,k)\wT^{\lambda\rho}(y+\ii z\beta(x))\rangle_{c,\beta},
\end{equation}
where the bracket $\langle\;\rangle_{c,\beta}$ denotes the connected correlator traced with the density matrix in absence of thermal vorticity
\begin{equation}
  \wrho_\beta = \frac{1}{Z_\beta} \exp \left[ -\wP\cdot\beta(x) - \zeta(x) \wQ \right].
\end{equation}
To obtain the spin vector (\ref{meanspf0}) we need the axial part of (\ref{eq:DeltaVortWig}), which, according to the definition (\ref{eq:WignFA}), is
\begin{equation}
\begin{split}
\Delta_\varpi \mathcal{A}_+^\mu(x,k) =& -\int\frac{\di^4 s}{(2\pi)^4}\int_0^1\di z\int_\Sigma \di\Sigma_\lambda(y)\varpi_{\rho\kappa}(y-x)^\kappa\e^{-\ii k\cdot s}\\
 &\times\langle \bar{\psi}^+(x+s/2)\gamma^\mu\gamma^5\psi^+(x-s/2)\wT^{\lambda\rho}(y+\ii z\beta(x))\rangle_{c,\beta} .
\end{split}
\end{equation}
Expanding the operators inside the correlators in terms of multi-particle states, see \cite{Buzzegoli:2021jeh}, the leading contribution of the Wigner function is given in terms of the matrix elements of the axial current
\begin{equation}
\label{eq:MEAxial}
\langle q',\tau|\wj_A^\mu(0)|q,\tau\rangle  = \bar{u}_\tau(q') A^\mu(q,q') u_\tau(q)
	= \bar{u}_\tau(q') \left[ F_{A1}\gamma^\mu\gamma^5 +F_{A2}\frac{(q'-q)^\mu}{2m}\gamma^5 \right] u_\tau(q),
\end{equation}
and of the energy momentum tensor:
\begin{equation}
\label{eq:MEEMT}
\langle q',\tau|\wT^{\mu\nu}(0)|q,\tau\rangle =\bar{u}_\tau(q') M^{\mu\nu}\left((q'+q)/2,q'-q\right) u_\tau(q).
\end{equation}
As the system is supposed to be in the hydrodynamic regime, we have a separation of scales between microscopic and macroscopic interactions.
In this approximation, the short-distance interactions are taken into account in the matrix elements (\ref{eq:MEAxial}) and (\ref{eq:MEEMT}) which take place at finite temperature.
The long-distance correlations are taken into account in the thermal expectation values of the creation and annihilation operators
$\langle \widehat{a}_{\tau'}^\dagger(q')\widehat{a}_\tau(q)\rangle =\delta_{\tau,\tau'}\delta^3(\vec{q}-\vec{q}')n_F(\beta\cdot q-\zeta)$.
One eventually obtains that
\begin{equation}
\label{eq:DelatA1}
\begin{split}
\Delta_\varpi \mathcal{A}_+^\mu(x,k) = & F_{A1}(k,0) \varpi_{\kappa\rho}u_\lambda \Delta^\kappa_{\;\kappa'}
	\frac{\theta(k_0)}{2\varepsilon_k}\frac{\delta(k^2-m^2)}{(2\pi)^3} n_F(k)(1-n_F(k))\\
& \times\left\{\ii \frac{\partial}{\partial p'_{\kappa'}}\tr\left[ M^{\lambda\rho}\left((p'+k)/2,p'-k\right)
	(\slashed{k}+m)\gamma^\mu\gamma^5(\slashed{p}'+m)\right]\right\}_{p'=k},
\end{split}
\end{equation}
where $\Delta^\kappa_{\kappa'}=\eta^\kappa_{\kappa'}-u^\kappa u_{\kappa'}$.
Finally, plugging the Eq. (\ref{eq:EMTinMedium}) in Eq. (\ref{eq:DelatA1}) and tracing gives the axial Wigner function induced by thermal vorticity:
\begin{equation}
\Delta_\varpi \mathcal{A}_+^\mu(x,k) = - g_\Omega(k,0)F_{A1}(k,0) \frac{\theta(k_0)\delta(k^2-m^2)}{(2\pi)^3} n_F(k)(1-n_F(k))\epsilon^{\mu\nu\sigma\tau}\varpi_{\nu\sigma}k_\tau .
\end{equation}
The spin polarization of a free field, using $g_\Omega(k,0)=1$, $F_{A1}(k,0)=1$, and applying the formula (\ref{meanspf0}), is the same as Eq. (\ref{eq:SpinPolFirstOrder}).
For an interacting field, the form factor $g_\Omega$ can receive radiative corrections, for instance like in Eq. (\ref{eq:AGMQED}), and the spin polarization induced by thermal vorticity can also receive radiative corrections.

\section{Summary and outlook}
I showed how spin polarization in a medium arises from the flow of the plasma and I discussed on how it depends on the properties of such medium.
I also showed how the spin polarization contains information about the gravitational form factor related to the spin-rotation coupling.
Therefore, measurements of the spin polarization in heavy-ion collisions provide a new tool to probe the properties of QCD matter at finite temperature.

In addition to probe the shear and vortical flow of the fluid and to test the dynamics of spin, several ideas on how to use spin polarization for investigating the QGP have already been proposed.
For instance, spin polarization can reveal local parity violations in QCD \cite{Du:2008zzb,Becattini:2020xbh}, it can probe the presence of the critical point \cite{Singh:2021yba}, and it can be used to estimate the energy loss of a jet traversing the medium \cite{Serenone:2021zef,Ribeiro:2023waz}.

\textbf{Acknowledgments.} M.B. thanks the U.S. Department of Energy for the support through the Grant No. DE-SC0023692.

\providecommand{\href}[2]{#2}\begingroup\raggedright\endgroup

\end{document}